# Abrupt Change Detection of Fault in Power System Using Independent Component Analysis

Harish C. Dubey, *Student Member, IEEE*
Department of Electronics & Communication Engineering
Motilal Nehru National Institute of Technology
Allahabad, INDIA
dubey.hc@rediffmail.com

Soumya R. Mohanty, Nand Kishore, *Member, IEEE*
Department of Electrical Engineering
Motilal Nehru National Institute of Technology
Allahabad, INDIA
soumyaigit@gmail.com; nand_research@yahoo.co.in

*Abstract*— This paper proposes a novel fault detector for digital relaying based on independent component analysis (ICA). The index for effective detection is derived from independent components of fault current. The proposed fault detector reduces the computational burden for real time applications and is therefore more accurate and robust as compared to other approaches. Further, a comparative assessment is carried out to establish the effectiveness of the proposed method as compared to the existing methods. This approach can be applied for fault classification and localization of a distance relay reflecting its consistency in all system changing conditions and thus validates its efficacy in the real time applications. The method is tested under a variety of fault and other disturbance conditions of typical power system.

*Keywords*— Abrupt change, blind source separation, digital relaying, distance relay, fault detection, independent component analysis, statistical signal processing.

## I. INTRODUCTION

Every power system is provided with a protective relay which ensures better performance while maintaining minimum disturbance and damage. In last few years, digital relays have replaced their solid-state-device counterparts due to their fast, accurate and reliable operation. The fault diagnosis unit of digital relays contains a fault detector (FD) unit in addition to fault classification and fault localization unit [1]-[2].

In recent years, a number of methods are available in the literature for detection of power system faults. Fault can be detected based on the comparison of difference between the value in current samples for two consecutive cycles being greater than threshold value and phasor comparison scheme [3]–[4]. However it has the limitation due to the difficulties in modeling the fault resistance. A Kalman filter–based approach [5]-[7] has been proposed in order to detect power system faults. Wavelet based approach [8] is used to detect the abrupt change in the signal. The synchronized segmentation is applied for disturbance recognition [9]. Then, application of adaptive whitening filter and wavelet transform has been used to detect the abrupt change in the signal [10]. However, these methods are sensitive to frequency deviation, presence of noise and harmonics.

In this paper, algorithm for abrupt change detection is proposed where the index for detection is derived from independent component of current samples. The proposed method performance have been tested under the presence of noise, harmonics and with frequency variation and found to be accurate. Independent component analysis (ICA) is selected for feature extraction because of its reliability to extract the relevant and useful features. Further, the proposed approach is compared with three existing approaches available in literatures. The first one of these is a detector based on comparison of sample value with one cycle. The second one being a differential approach based on phasor estimation [3] while third is a moving-sum based detector where sum over one cycle of faulty current samples is chosen as index for detection [11].

Rest of the paper is arranged as follows; section II gives a brief description of three approaches used for comparative assessment of proposed approach followed by section III, which gives a brief description of independent component analysis technique. Next, section IV presents the discussion on the proposed approach based on independent component analysis while section V presents the testing of the proposed approach. Finally, conclusions are given in section VI.

## II. FAULT DETECTION TECHNIQUES USED FOR POWER SYSTEM BASED ON TIME-SERIES DATA

This section gives brief description of fault detection techniques used for power system based on time-series data. These three techniques are used to carry out the comparative assessment of proposed approach in changing conditions of the system. All these approaches are based on deterministic modeling of faulty current signal obtained from a typical power system.

### A. Sample Comparison (SC)

The first approach for fault detection is the conventional. Here, decision is taken out by computing the difference of current sample of signal with corresponding sample of the one cycle earlier. Under normal conditions, the computed difference comes out to be zero. When there is a fault in the system, the current signal gets distorted and consequently computed difference become significant. If the computed difference remains greater than a threshold value for three consecutive samples, a fault is reported by the FD unit. Let the discrete current signal be

$$i(k) = I_m \sin(k\Omega + \phi) \tag{1}$$

Where, $I_m$ is the peak of the signal, $\Omega$ is the discrete angular frequency and $\phi$ is the phase angle. Then the index is derived as follows:

$$i_{change}(k) = i(k) - i(k-N) \qquad (2)$$
$$Index_{sc}(k) = |i_{change}(k)| \qquad (3)$$

Where, $k$ is the time-instant and $N$ is the window size of one period. If

$$Index_{sc}(k) > (Index_{sc})_{threshold}, \qquad (4)$$

A fault is reported by FD unit.

*B. Phasor Comparison (PC)*

This approach for fault detection is based on estimation of the phasor [3]. It is a relatively fast algorithm based on the derivative of the current signal. If the discrete current signal is,

$$i(k) = I_m \sin(k\Omega + \phi) \qquad (5)$$

Where, $I_m$ is the peak of the signal, $\Omega$ is the discrete angular frequency and $\phi$ is the phase angle. Then at any instant, $k$, the peak-value of the signal can be estimated as,

$$\left(\hat{I}_m(k)\right)^2 = \left(\frac{i''(k)}{\Omega^2}\right)^2 + \left(\frac{i'(k)}{\Omega}\right)^2 \qquad (6)$$

Where $\hat{I}_m(k)$ is the peak estimate of the signal, and $i'(k)$ and $i''(k)$ are the first and second derivatives of discrete current signal respectively. The peak estimate is the magnitude of fundamental phasor at k-th estimate. The magnitude of the current phasor obtained at *k-th* instant is compared with that at (*k*-3)-th instant. If the difference is more than the threshold value for three successive samples, a fault is reported by FD. The derivation of index is as follows:

$$I_{change}(k) = \hat{I}_m(k) - \hat{I}_m(k-3) \qquad (7)$$
$$Index_{pc}(k) = |I_{change}(k)|, \qquad (8)$$

If

$$Index_{pc}(k) > (Index_{pc})_{threshold} \qquad (9)$$

Then FD detects the fault. As the method is derivative based, it is found to be sensitive to noise and signal distortions.

*C. One-Cycle-Moving-Sum (OCMS):*

This approach involves the computation of one cycle sum of current samples obtained from the power system [11]. This approach is based on the symmetrical nature of the current waveforms in power system. In absence of fault, the computed sum comes out to be zero. However, on occurrence of fault in the power system, the corresponding sum will be non-zero or equivalently greater than a chosen threshold. For on-line implementation, once a new sample is obtained, the oldest sample is discarded and the sum is recalculated for the new window. Thus, only one addition and one subtraction is required at each step of computation. For large variation in system conditions such as frequency variations, the window size needs to be made adaptive to generate the zero sums in normal condition. However, such variations are not common in large power system. Let the discrete current signal be

$$i(k) = I_m \sin(k\Omega + \phi) \qquad (10)$$

Then, the derivation of index is as follows:

$$i_{sum}(k) = \sum_{l=k-N+1}^{k} i_l \qquad (11)$$

Where, $N$ is the window size for one cycle.

$$Index_{ocms}(k) = |i_{sum}| \qquad (12)$$

If $Index_{ocms}(k) > (Index_{ocms})_{threshold} \qquad (13)$

A fault is reported by FD unit. Also,

$$i_{sum}(k) = i_{sum}(k-1) + i(k) - i(k-N) \qquad (14)$$

The above eqn. (14) shows that, for on-line computation, only one addition and one subtraction is required at each step.

## III. INDEPENDENT COMPONENT ANALYSIS

Since ICA is based on the statistical properties of signals, it works accurately in non-deterministic modeling of the signals [12]. For ICA to be applied, following assumptions for the mixing and demixing models needs to be satisfied:
1. The source signals $s(t_i)$ is statistically independent.
2. At most one of the source signals is Gaussian distributed.
3. The number of observations $M$ is greater or equal to the number of sources $N$ ($M \geq N$).

In addition to blind separation of sources, ICA is also used for representing data as linear combination of latent variables. There are different approaches for estimating the ICA model which are based on the statistical properties of signals. Some of the methods used for ICA estimation are:
1. by maximization of nongaussianity
2. by minimization of mutual information
3. by maximum likelihood estimation,
4. by tensorial methods

Blind source separation algorithm estimates the source signals from observed mixtures. The word 'blind' emphasizes that the source signals and the way the sources are mixed, i.e. the mixing model parameters, are unknown or known very imprecisely. Independent component analysis is a blind source separation (BSS) algorithm, which transforms the observed signals into mutually statistically independent signals. The ICA algorithm has many technical applications including signal processing, brain imaging, telecommunications and audio signal separation [12] – [14].

*A. ICA estimation by maximization of nongaussianity:*

A measure of nongaussianity is negentropy $J(y)$ which is the normalized differential entropy. By maximizing the negentropy, the mutual information of the sources is minimized. Also, mutual information is a measure of the independence of random variables. Negentropy is always non-negative and zero for Gaussian variables. [12]

$$J(y) = H(y_{gauss}) - H(y) \qquad (15)$$

The differential entropy $H$ of a random vector $y$ with density $p_y(\eta)$ is defined as

$$H(y) = -\int p_y(\eta) \log p_y(\eta) d\eta \qquad (16)$$

In equation (15) and (16), the estimation of negentropy requires the estimation of probability functions of source signals which are unknown. Instead, the following approximation of negentropy is used:

$$J(y_i) = J\left(E\left(w_i^T x\right)\right) = \left[E\left\{G\left(w_i^T x\right)\right\} - E\left\{G\left(y_{gauss}\right)\right\}\right]^2 \quad (17)$$

Here, $E$ denotes the statistical expectation and G is chosen as non-quadratic. Assuming that we observe $n$ linear mixtures $x_1, ..., x_n$ of $n$ independent components:

$$x_j = a_1 s_1 + a_2 s_2 + .... + a_n s_n \text{ For all } j \quad (18)$$

We assume that each mixture $x_j$ as well as each independent component $s_k$ is a random variable, instead of a time dependent signal. Without loss of generality, we can assume that both the mixture variables and the independent components have zero mean. If this is not true, then the observable variables $x_j$ can always be centered by subtracting the sample mean, which makes the model zero mean. It is convenient to use vector-matrix notation instead of the sums like in the previous equation. Let us denote by **x,** the random vector whose elements are the mixtures $x_1, ..., x_n$ and likewise by **s** the random vector with elements $s_1, ..., s_n$. Let us denote by **A** the matrix with elements $a_{ij}$. All vectors are taken as column vectors; thus $x^T$, or the transpose of **x**, is a row vector. With this vector-matrix notation, the above mixing model becomes:

$$x = As \quad (19)$$

Denoting the column of matrix **A** by $a_j$ the model can also be written as

$$x = \sum_{i=1}^{n} a_i s_i \quad (20)$$

The statistical model in eqn (20) is called independent component analysis, or ICA model. The ICA model is a generative model and the independent components are latent variables, meaning that they cannot be directly observed. Also the mixing matrix is assumed to be unknown. All we observe is the random vector **x**, and we must estimate both $A$ and $s$ using it. The starting point for ICA is the very simple assumption that the components $s_i$ are statistically independent. We also assume that the independent components have non-Gaussian distributions. Then, after estimating the matrix **A**, we can compute its inverse, say **W**, and obtain the independent component simply by:

$$s = Wx \quad (21)$$

Fast ICA is an efficient algorithm based on fixed-point iteration used for estimation of ICs in time series data [15]. This approach for ICs estimation is 10-100 times faster than the other methods that are used to reduce data dimension.

## IV. PROPOSED FAULT DETECTION METHOD

This section presents the algorithm of the proposed method for detection of abrupt changes due to occurrence of fault in the power system. An abrupt change detector based on independent components of current samples is proposed in this section. The index for detection is derived from independent components of current sample obtained from data acquisition system.

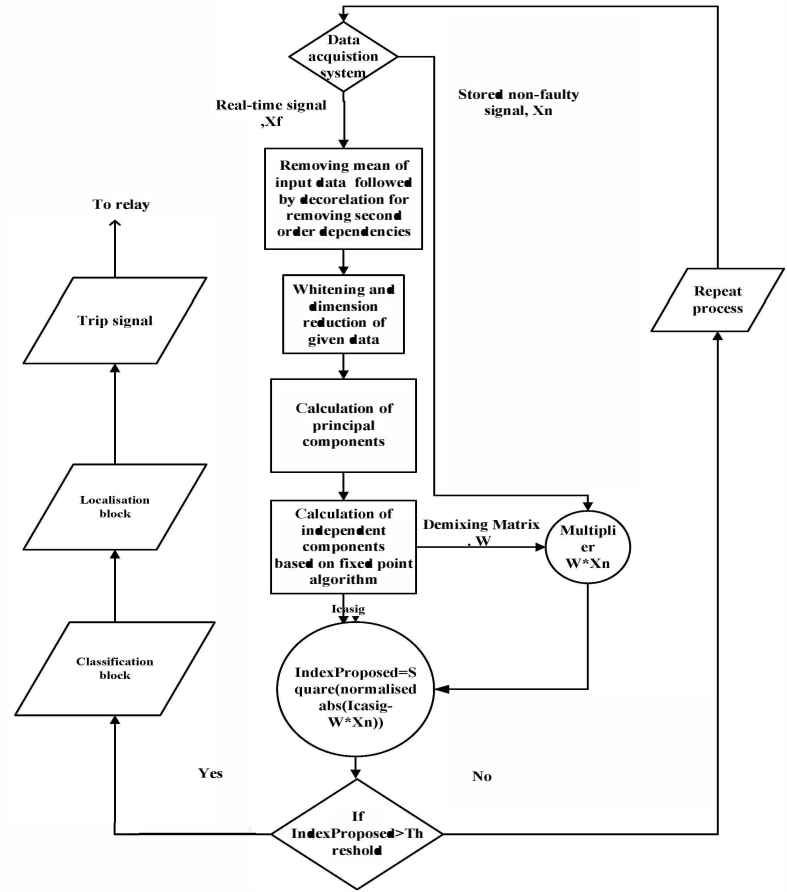

Fig.1 Flowchart of the proposed algorithm.

The proposed algorithm has been tested on simulation data and is explained below:

(i) Data has been obtained from MATLAB/Simulink model of the interconnected power system considered in this work. Also in this study, the pre-fault signal can be taken as non-faulty signal. The signal is first passed through the detection block followed by classification block and finally through localization block for deciding logic for trip signal system. This constitutes fault diagnosis system.

(ii) The simulated signal is passed through first block where the removal of mean and de-correlation (for removal of second order dependencies) is done. This constitutes the first level of pre-processing. The output of this block is fed to the next block.

(iii) In this block, whitening of data followed by dimension reduction is performed for reducing redundancy in data. Output of this block is fed to third block.

(iv) Now, principal components (PC) of data are determined and fed to next block.

(v) Here, independent components of data are calculated using fixed point iteration of Fast ICA algorithm [12], [15].

(vi) The ICA block returns demixing or separating matrix, $W_f$ along with independent component, $s_f$ of real time signal. For calculation of these variables matrix, $x_f$ is constructed from real time signal samples.

(vii) The stored signals or the pre-fault signals are used to construct matrix, $x_n$ for the derivation of index.

The index is derived as

$$Index_{proposed}(k) = (normalised(abs(W_f(k)*x_n(k) - s_f(k))))^2) \quad (22)$$

The fault is detected when $Index_{proposed}(k)$ is greater than a certain threshold $(Index_{proposed})_{threshold}$. The threshold, $(Index_{proposed})_{threshold}$ is evaluated by decision block and appropriate actions are taken. This information is then passed to classification block. The flowchart of the proposed algorithm is illustrated in Fig. 1.

## V. COMPARATIVE ASSESSMENT AND TESTING OF THE PROPOSED ALGORITHM

A three phase transmission line (200km, 230 kV, 50 Hz) connecting two systems with MOV and series capacitor kept at the middle of line as shown in Fig. 2 has been considered for comparative assessment of the performance between existing and proposed algorithms. We have demonstrated the comparative assessment of the performance of the various algorithms by considering the fault at the same instant at different conditions for the sake of better clarity of the result. The typical power system model in MATLAB/Simulink is used in obtaining simulation data. At the receiving end, the combination of linear and non-linear load is used. Depending on the switching of non-linear load, harmonics are obtained in the current signal. The testing data is obtained through simulation of considered power system under different system changing conditions. A sampling rate of 1 kHz and a full cycle window of N= 20 (50 Hz nominal frequency) has been chosen for testing.

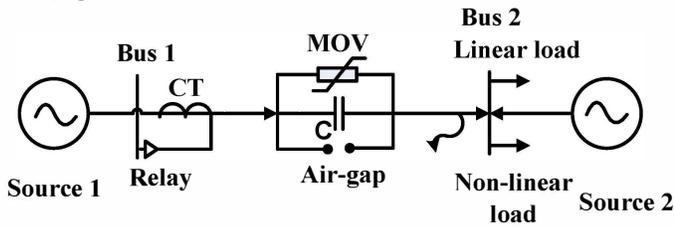

Fig.2 A 230kV, 50 Hz Power System.

Faults of various types are simulated at different locations and the performance of the algorithms is assessed. To demonstrate the potential of the approach only few cases of fault occurrence towards the farther end of the line are demonstrated here. Nevertheless, the proposed method responds similarly to other types of power system faults too. Single line-to-ground faults (AG-type) at 80% of the line have been created at different inception angles and the corresponding phasor current, tapped at Bus 2 is processed through the different algorithms and the detection indices are computed and normalized for comparison. As the FD is expected to be fast enough to detect the inception of fault within few milliseconds, first few sampling periods are important to adjudge the performance of the algorithm.

TABLE 1 PARAMETERS OF THE POWER SYSTEM MODEL

| System voltage | 230 kV |
|---|---|
| System frequency | 50 Hz |
| Voltage of source 1 | 1.0 0 degree pu |
| Voltage of source 2 | 1.0 ∠ 10 degree pu |
| Transmission line length | 200 km |
| Positive sequence | R = 0.0321 (ohm/km), L = 0.57(mH/km), C = 0.021 (uF/km) |
| Zero sequence | R = 0.0321 (ohm/km), L = 1.711 (mH/km), C = 0.021 (uF/km) |
| Series compensated | 70 % C =176.34 uF |
| MOV Vref | 40 kV 5 MJ |
| Current transformer (CT) | 230kV, 50 Hz, 2000:1(turns ratio) |

### A. Abrupt change detection without noise

The interconnected power system as shown in the Fig. 2 is simulated in MATLAB/Simulink. A L-G fault has been created at 0.065 s with the system frequency as 50 Hz. Comparative assessment of proposed algorithm is carried out with existing algorithms and shown in Fig. 3. Here, all the indices approximately indicates the fault situation with minimum (1-2) sample delay after the inception of the fault. In the post-fault region, almost all the algorithm exhibits consistent results.

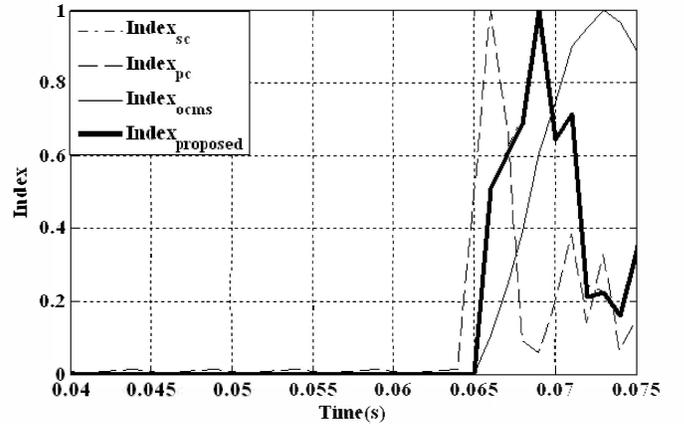

Fig.3 Fault detection without noise.

### B. Abrupt change detection with noise

A phase-to-ground fault is created at 0.065 s with the system operating at nominal frequency of 50 Hz and a noise signal of 20 dB SNR added to original one for performance assessment. The normalized indices are shown in Fig. 4. It is observed that values of indices determined from three algorithms as discussed in section II are significant even before the occurrence of fault inception. However, index of proposed method demonstrates the instant of fault inception correctly. $Index_{sc}$ as crosses the threshold before the occurrence of fault i.e. in the steady state situation may be mis-interpreted as the fault even if there is no fault. *Index$_{ocms}$* also exhibits a non-zero variation although it sums the total noisy signal over a fixed data window i.e. 20 samples per cycle. Thus, the indices exhibits variation in pre-fault region and are not consistent in the post-fault region as well. On contrary, the proposed method shows almost zero index value in pre-fault region and consistent index in the post-fault region also.

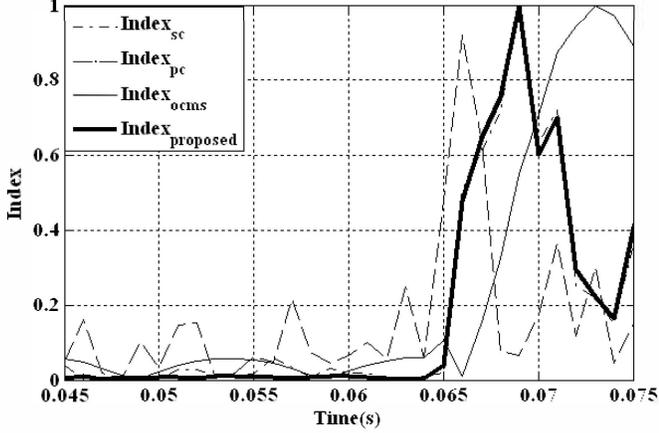

Fig.4. Fault detection with 20 dB noise.

### C. Abrupt change detection with DC offset

A phase-to-ground fault has been created at 0.065 s in the system. The current signal is processed through the different methods as described in Section II and IV. The normalized indices are given in Fig. 5. It is observed that the existing methods fail to sense the change immediately. In presence of DC offset, algorithm based on accumulated sum of samples in fixed data window, i.e. moving sum deviates from nominal value of zero or very small value of the threshold. As a matter of fact, before the occurrence of fault, moving sum algorithm is not suitable approach for the description of the fault occurrence. Similarly, the other existing algorithm also performs poorly in the pre-fault period. On the other hand, the proposed approach does not deviate from nominal threshold in the pre-fault region. Even if the algorithm based on PC approach shows comparable performance but still it is inconsistent in post fault region since its value becomes equal to threshold value. This is misinterpreted as fault inception.

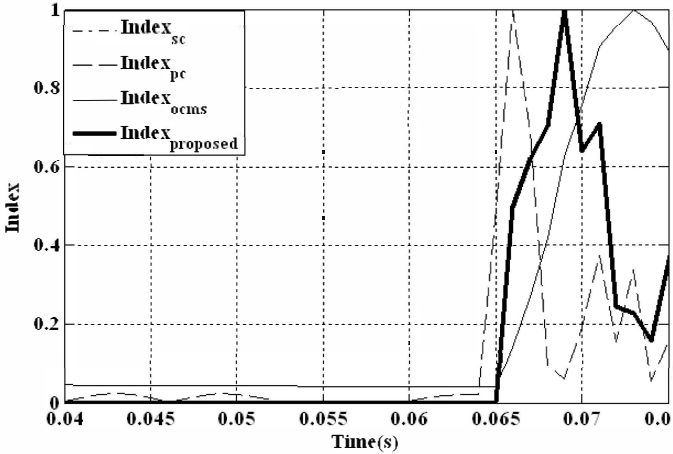

Fig.5 Fault detection in presence of dc-offset.

### D. Abrupt change detection with harmonics

With the incorporation of non-linear load at Bus 2, the harmonics are generated in addition to the fundamental components in the signal. A phase-to-ground fault has been created at 0.065 s in the system. The current signal is processed through the different methods as described in earlier sections. The normalized indices have been plotted in Fig. 6. A favorable detection of fault by proposed algorithm over existing ones is observed.

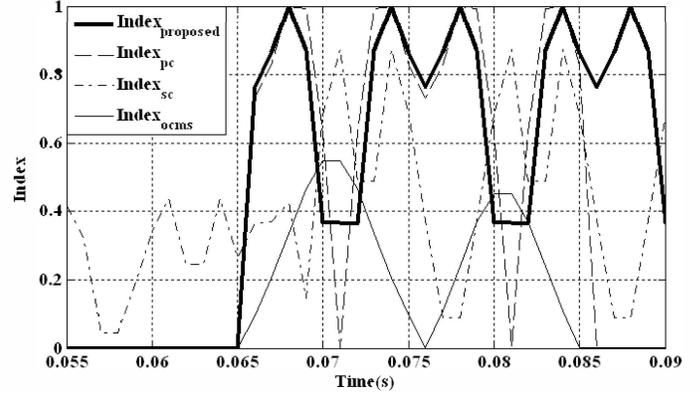

Fig. 6 Fault detection in presence of harmonics.

### E. Abrupt change detection with change in frequency

Frequency variations are common in power systems. Thus, the frequency estimation is indispensable for demonstrating the performance of the existing algorithms such as sample comparison, phasor approach, moving-sum approach etc. In the study, firstly the frequency is estimated by variable leaky-least mean square (VL-LMS) that tracks the original frequency change faster than complex LMS algorithm [16]-[20]. After frequency estimation, assessment of the existing algorithm is demonstrated. A phase-to-ground fault has been created at 0.065 s in the system operating at nominal frequency of 52 Hz. The normalized indices have been plotted in Fig.7. As indicated, the proposed approach is still consistent against the indices of existing algorithms in tracking the point of change.

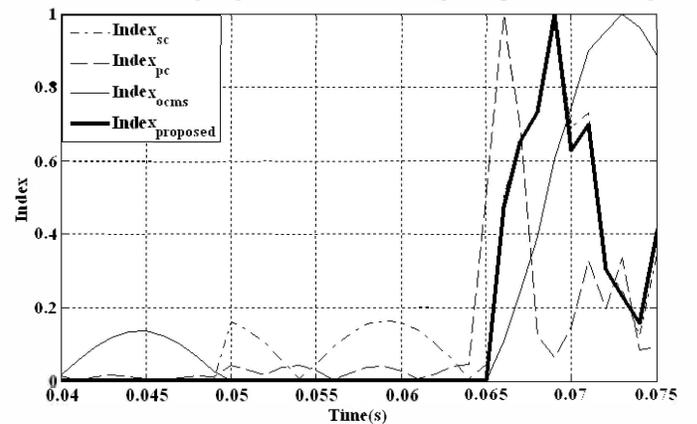

Fig.7 Fault detection with change in frequency.

## VI. Conclusion

Fault detection for relaying application is a challenging task in the presence of noise, harmonics and frequency change of signal. Traditional methods are based on deterministic modeling i.e. sinusoidal behavior of current/ voltage and are therefore sensitive to noise. In this paper, a novel fault detection algorithm was proposed based on the independent components of current signal. The proposed technique does not assume sinusoidal behavior of current/ voltage signal. The performance of the method was assessed through simulation with different fault data and compared with existing techniques. It has been found that this method provides very consistent results under all the fault conditions. The method was compatible with any sampling frequency conventionally being used for relaying applications.

## Acknowledgment

Authors would like to thank Prakash K. Ray, Research Scholar, Electrical Engineering Department, MNNIT, Allahabad for helpful discussions.